# GAMMA RAY BURSTS FROM MINIJETS

*Nir J. Shaviv* [1] *& Arnon Dar*[2]
Physics Department, Technion – Israel Institute of Technology
Haifa 32000, Israel

## ABSTRACT

Striking similarities exist between high energy gamma ray emission from active galactic nuclei (AGN) and gamma ray bursts (GRBs). They suggest that GRBs are generated by inverse Compton scattering from highly relativistic electrons in transient jets. Such jets may be produced along the axis of an accretion disk formed around stellar black holes (BH) or neutron stars (NS) in BH-NS and NS-NS mergers and in accretion induced collapse of magnetized white dwarfs (WD) or neutron stars in close binary systems. Such events may produce the cosmological GRBs. Transient jets formed by single old magnetized neutron stars in an extended Galactic halo may produce a local population of GRBs. Here we show that jet production of GRBs by inverse Compton scattering can explain quite simply the striking correlations that exist between various temporal features of GRBs, their duration histogram, the power spectrum of their complex multipeak light curves, their power-law high energy spectra and other features of GRBs. Some additional predictions are made including the expected polarization of gamma-rays in the bursts.

---

[1]E-mail: nir@phastro.technion.ac.il

[2]E-mail: PHR19AD@technion.technion.ac.il

## 1. Introduction

The Burst and Transient Source Experiment (BATSE) on board the Compton Gamma Ray Observatory (CGRO) has established that the distribution of gamma ray bursts (GRBs) is isotropic but bound in radial direction (Fishman et al. 1991; Meegan et al. 1992; 1994). This suggests that GRBs are either cosmological (e.g., Fenimore et al. 1992; Paczynski 1992; Mao and Paczynski 1992; Piran 1992; Dermer 1992; Kouveliotou et al. 1992; Paciesas et al. 1992; Norris et al. 1992) and/or originate in an extended halo of our Milky Way galaxy (e.g. Brainerd 1992; Eichler and Silk 1992; Hartmann 1992; Li and Dermer 1992). The implied luminosities and the observed variability of the GRBs on time scales as short as 0.2 ms (Bahat et al. 1992) suggest a neutron star (NS) or a black hole (BH) origin. The leading candidates for gamma ray bursters in an extended galactic halo are single old magnetized NS (e.g., Fishman et al. 1978; Jennings and White 1980; Jennings 1984 Shklovski and Mitrofanov 1985; Atteia and Hurley 1986; Yamagami and Nishimura 1986). Indeed, the observed high birth velocities of radio pulsars with an estimated mean velocity of $450 \pm 90\ km\ s^{-1}$ (e.g., Lyne and Lorimer 1994) imply a very large spherical halo of old neutron stars that is consistent with the BATSE observations (Li and Dermer 1992; Eichler and Silk 1992; Lyne and Lorimer 1994). A phase transition in the NS or a starquake may generate a GRB (see e.g., Higdon and Lingenfelter 1990 and references therein) although neither a realistic mechanism for the efficient conversion of the gravitational energy release into a GRB has been proposed, nor other characteristic features of GRBs (short time scale variability, complexity and bewildering variety of light curves, durations that spans six orders of magnitude, and a simple power-law spectrum at high energies) have been explained.

Accretion induced collapse of a white dwarf (WD) or a neutron star, or mergers in BH - NS, NS - NS close binary systems due to gravitational wave emission, with conversion of gravitational binding energy to a relativistic expanding fireball via the $\nu + \bar{\nu} \to e^+ + e^-$ process, were proposed as the origin of cosmological GRBs (Goodman, Dar and Nussinov 1987, Eichler et al. 1989, Ramaty et al. 1990; Paczynski 1991, Dar et al. 1992). Although NS - NS mergers may explain the event rate of GRBs (Narayan et al. 1991, Phinney 1991) and their intensity distribution (e.g., Dermer 1992, Mao and Paczynski 1992) they do not provide a simple explanation for the above characteristics of GRBs and for a duration histogram that shows a bimodality structure and span six orders of magnitude. However, the absence of simple explanations of these features does not mean that such explanations cannot be found. In fact, potential explanations of these features have been suggested (e.g., Dar and Nussinov 1992) and should be further investigated: The time variability may reflect variation in the optical thickness along the line of sight to the GRB due to obscuring material (wind, ejected shell or accreted material) that orbits the burster and is Rayleigh-Taylor unstable to the radiation pressure of the GRB. It may also be due to the complex interaction between the relativistic expanding fireball and interstellar matter (Rees and Meszaros 1992; Meszaros and Rees 1993). The bimodality of the duration histogram may reflect the different natures of the fireballs generated by BH (shorter and weaker due to strong gravitational redshift effects) and by NS (longer and stronger), while the total duration may depend on the baryon content of the fireball and the surrounding material (Ramaty et al. 1990; Dar et al. 1992). Finally, the power-law spectrum of the high energy photons may be produced naturally by the expanding fireball.

However, in this paper we would like to explore an alternative universal mechanism for production of GRBs, namely, inverse Compton scattering of photons from transient jets of highly relativistic particles. Such transient jets may be produced by old magnetized NS in the extended NS halo of our Milky Way galaxy and give rise to the population of short GRBs. (Jets of highly relativistic particles produced by magnetized rotating NS are the most solidly established point sources of continuous high energy gamma ray emission in the galaxy). Transient jets of highly relativistic particles may also be produced in the birth of black holes and neutron stars in external galaxies in a mechanism similar to that by which active galactic nuclei (AGN) produce their jets of highly relativistic particles. In fact, the Energetic Gamma Ray Experiment Telescope (EGRET) on the Compton Gamma Ray Observatory (CGRO) has detected 31 AGN in high energy gamma rays thus far (Thompson 1993; Hunter 1994). Except for the different time scales, striking similarities exist between the gamma ray bursts observed by the BATSE and EGRET telescopes and the high energy gamma ray emission from the EGRET AGN sources:

1. Both show a rapid variation of luminosity as function of time, a diversity of light curves and a complex time structure with exponential temporal autocorrelation functions (see below).

2. Both show a power-law spectrum, $dN_\gamma/dE \sim E^{-\alpha}$ with approximately the same powers, $1.5 \lesssim \alpha \lesssim 2.5$ , that extends to high energies.

3. The emission in other bands of the electromagnetic spectrum by gamma ray bursters is either directed away from us or must have a much lower luminosity, since no emission other than gamma rays has been detected thus far from the gamma ray bursters. A similar feature exists for the AGN that were detected by EGRET. Their apparent luminosity in high energy gamma rays is typically larger by more than two orders of magnitude than their observed luminosities in other bands of the electromagnetic spectrum.

4. If GRBs, like AGN, are at cosmological distances then their peak luminosities, within the solid angle which their emission substands, are highly super-Eddington for any compact object which can produce their short time variabilities.

5. Both GRB and AGN gamma ray spectra do not show a cutoff due to self absorption via the pair production process $\gamma + \gamma \to e^+ + e^-$ (e.g., Schaefer et al. 1992).

All the EGRET AGN sources are also radio-loud, flat radio spectrum sources. The exclusive association of EGRET AGN sources with radio-loud flat radio spectrum sources strongly suggests that the observed gamma rays are produced by jets that are beamed in our direction, in a similar fashion to the radio emission. This picture fits well into the unified picture of AGN being an accretion disk, surrounded by a thick torus of cooler gas, feeding a massive black hole and emitting a jet of highly relativistic particles along the axis of the disk. This picture is supported by the observation of radio-frequency jets, superluminal motion in some sources, super Eddington luminosity within the solid angle substended by the emission and other observations (for a review see Osterbrock 1991). According to this picture all AGN are different aspects of the same phenomena viewed at different angles in the sky and/or at different epochs (in the case of OVV quasars the jet is aimed almost directly at the observer while in Seyfert 1 and 2 galaxies the line of sight is oriented at a substantial angle with respect to the axis of the disk and torus). Additional support for the beaming of gamma-ray emission from AGN comes from the fact that several of the gamma-ray loud AGN have large redshifts ($z > 1$) and are emitting gamma rays at an extra ordinary rate ($\geq 10^{49} b\ erg\ s^{-1}$ where b is a beaming factor) while at the same time many active galaxies that are relatively close to Earth, including some blazars, have not been detected in high energy gamma-rays. Moreover, in view of the continuing rapid variations in luminosity observed on timescales $\Delta t \lesssim 1h$, the beaming of gamma rays is required in order to avoid a large induced optical thickness due to the pair production process, $\gamma + \gamma \to e^{+} + e^{-}$, for gamma rays that are emitted from a region whose dimensions are restricted by causality to satisfy $R < c\Delta t/(1+z)$ . Finally, the power-law spectrum of the emitted high energy gamma rays is also consistent with a jet origin of the high energy gamma rays.

The mechanism by which an accretion disk around a massive black hole produces a relativistic jet of particles is not clear yet and may involve a complex mixture of plasma physics, hydrodynamics, electrodynamics and general relativity. Rather than trying to generalize theoretical models of relativistic jet formation by AGN to accreting neutron stars (NS) and stellar black holes (BH), we will assume that transient jets of highly relativistic particles are also produced by the accretion disk formed during the last stage of mergers caused by energy loss due to gravitational wave emission in close BH - NS and NS - NS binary systems, by accretion induced collapse of highly magnetized WD or NS and by starquakes in old magnetized NS. We will show that inverse Compton scattering of soft photons emitted by the accretion disk and/or by the thick torus of matter feeding the accretion disk, and/or by the jet itself (e.g., synchrotron self emitted radiation) can explain many of the observed features of gamma ray bursts including their duration histogram and the striking correlations between their temporal characteristics. We will also explain some other features of GRBs and predict the polarization of the gamma rays in the bursts. For the sake of simplicity and clarity we will use the simplest jet model.

Our paper is organized as follows: In chapter 2 we present observational evidence for some striking correlations between temporal characteristics of GRBs. In chapter 3 we present our jet model for production of GRBs and use it to predict these correlations. In chapter 4 we compare other predictions of the jet model with observations after including cosmological effects and threshold biasing effects. Additional predictions for future tests are made in chapter 5. A short discussion and conclusions are included in chapter 6.

## 2. Correlations and Temporal Features of GRBs

Recently Shaviv (1994) has found some striking correlations between various temporal characteristics of GRBs in the BATSE 1B catalogue (Fishman et al 1994):

1. The two different duration measures of GRBs adopted by BATSE, $T90$ and $T50$, are linearly correlated, where $T90$ is the time interval between 5% and 95% of the integrated number of gamma ray counts in the GRB and $T50$ is the time interval between 25% and 75% of the counts, respectively. This is demonstrated in Fig. 1a where $T90$ as function of $T50$ is plotted on a log-log plot for the GRBs in the BATSE 1B catalogue. As can be seen from Fig.1a the points seem to lie along a straight line with a slope of $0.99 \pm 0.015$, indicating a linear correlation between $T90$ and $T50$. Although, at first sight, this linear correlation may seem trivial, it is not. This is demonstrated in Fig.2 where we show $T90/T50$ versus $T50$ on a log-log plot for supernova explosions, with the same intrinsic luminosities and exponentially decaying light curve, which are uniformly distributed in a Euclidean space.

2. The time correlation length $\tau_0$ defined through the auto correlation function $A(\tau)$ (Link et al. 1994),

$$A(\tau) \equiv \frac{\langle [s(t+\tau) - b][s(t) - b] \rangle_{\mathrm{t}}}{\left\langle [\mathrm{s}(t) - b]^2 \right\rangle_t} \approx \mathrm{e}^{-\tau/\tau_0} \tag{1}$$

   with $s$ and $b$ being the signal and background, and the duration (T90) are also linearly correlated. This is demonstrated in Fig. 1b where we have plotted $\tau_0$ as function of $T90$ on a log-log plot for the GRBs analyzed by Link et al. (1993). (Link et al. obtained $\tau_0$ from measuring the area under $A(\tau)$ which is equivalent to measuring the slope of $\log A(\tau)$). As can be seen from Fig.1b the GRBs lie along a straight line whose slope is $0.92 \pm 0.07$ suggesting a linear correlation between $\tau_0$ and $T90$.

The above linear correlations can exist if either the average burst shape is linear in time (which is not the case) or the temporal behavior of GRBs is scale invariant, i.e., there is some mechanism that stretches the durations of the original bursts by a different factor for each burst. The origin of this scale invariant behavior can be either intrinsic or extrinsic. Examples of mechanisms which can cause such a behavior are cosmological time dilation, gravitational time dilation and geometrical effects.

A cosmological time dilation origin of the scale invariant behavior is unlikely because the durations of GRBs span at least six orders of magnitude while GRBs, if they are cosmological, cut off around a redshift of $z \approx 1$, i.e., the cosmological time dilation factors are typically less than $1 + z \approx 2$.

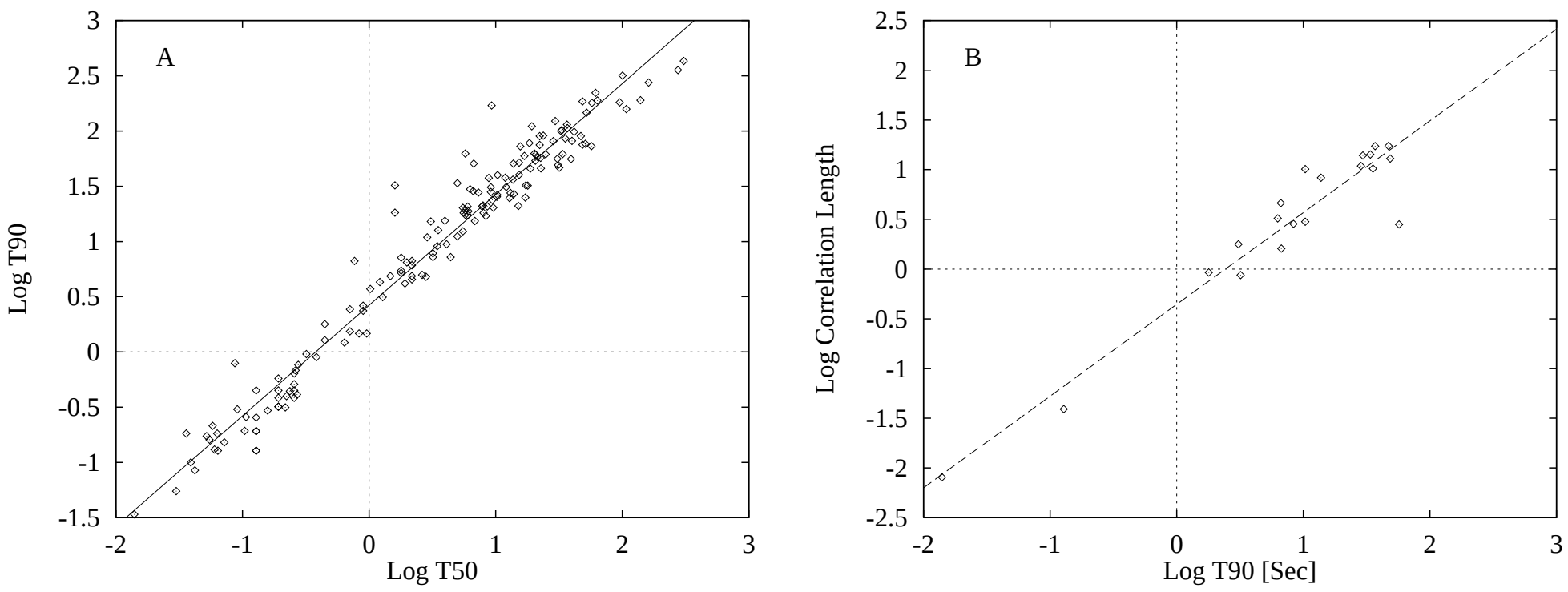

Fig. 1.— (a) A logarithmic plot of the $T90$ duration as function of the $T50$ duration for GRBs from the BATSE B1 catalogue. The straight line is a best fitted line and has a slope of $0.99 \pm 0.015$. (b) A logarithmic plot of the time correlation length $\tau_0$ as function of the $T50$ duration for the GRBs that were analyzed by Link et al. (1993). The straight line is a best fitted line and has a slope of $0.92 \pm 0.07$ .

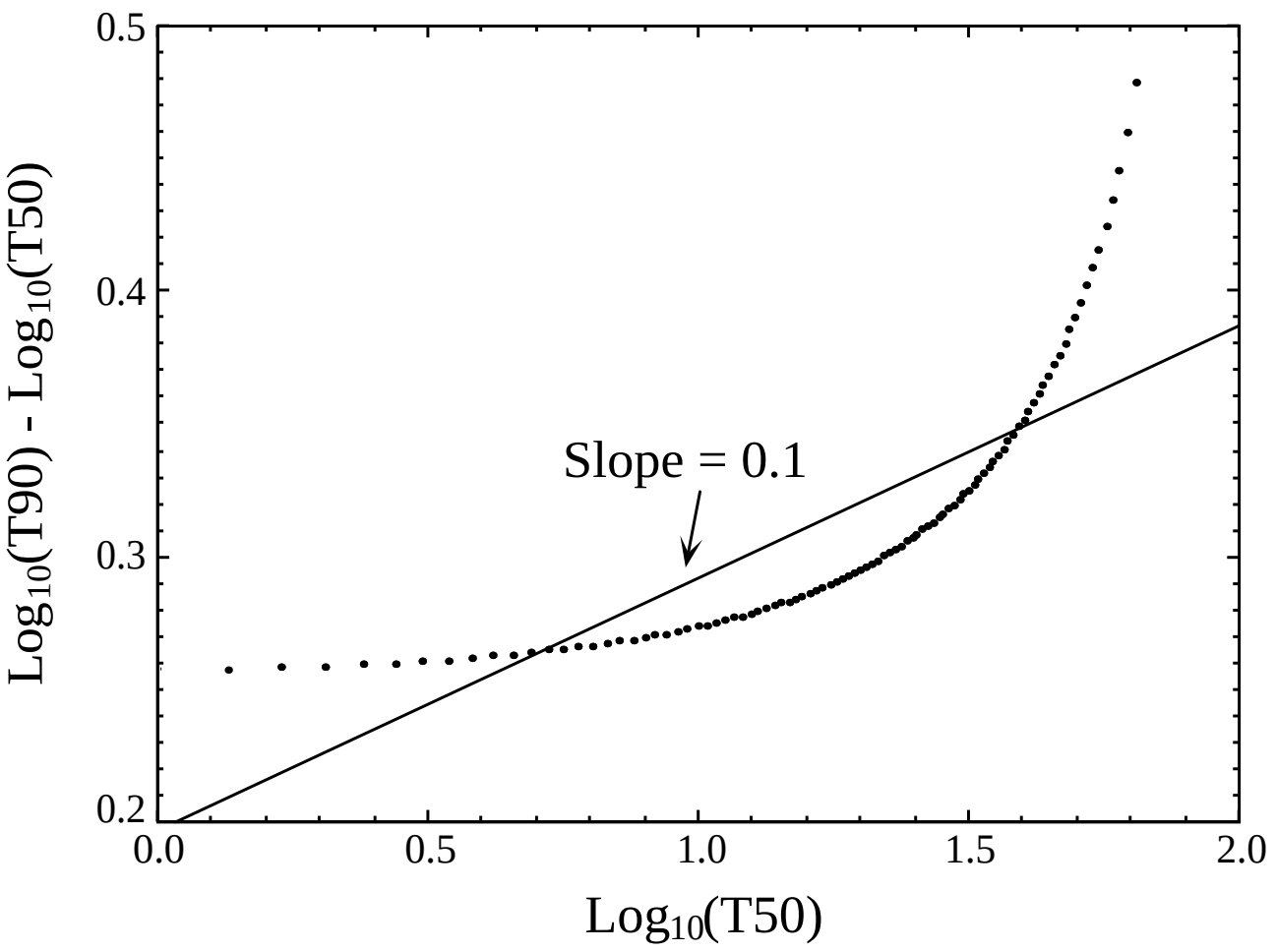

Fig. 2.— The durations ratio $T90/T50$ as function of $T50$ on a log-log plot for supernova explosions, with the same intrinsic luminosities and exponentially decaying light curves, which are uniformly distributed in a Euclidean space.

A gravitational redshift which is able to cause the observed wide spread of durations will also redshift the luminosities of the bursts by the same amount. Such a strong correlation is not observed in the data.

However, the origin of the above correlations can be geometrical. This will be shown in the next chapter, for GRBs which are produced by inverse Compton scattering of light from jets. (Although probably other explanations for the time stretching of the light curves of GRBs are possible, ultimately, a correct theory of GRBs should explain the above striking correlations).

In addition, it was found that the power spectra $P_i(\omega)$ of GRBs with complex light curves, $L_i(t)$, decrease like $\omega^{-2}$ for large values of $\omega$ ($\omega \gg 1/T90$) and, in fact, are well represented by

$$P_i(\omega) \sim \frac{1}{\omega^2 + \lambda_i^2} \tag{2}$$

with $\lambda_i \lesssim 1/T90$, where

$$P_i(\omega) = \left| \frac{1}{2\pi} \int_{-\infty}^{\infty} e^{i\omega t} L_i(t) dt \right|^2 . \tag{3}$$

This behavior is demonstrated in Figs. 3a-3h for arbitrarily selected GRBs from the BATSE 1B catalogue.

## 3. The Jet Model

Let us assume that gamma ray bursters form a highly collimated electron jet with a large Lorentz factor, $\gamma \gg 1$, and that the jet is illuminated for a short period of time by soft photons which are Compton scattered into different directions. Photons with an incident angle much larger than $1/\gamma$ are boosted by inverse Compton scattering to gamma ray energies. The model is described schematically in Fig.4.

### 3.1. Observed Fluences

For simplicity, we consider a jet with a fixed length $l_0$ emergent along the axis of a disk which illuminates it from behind by a short burst ($\Delta T \ll l_0/c$) of soft photons (see Fig. 4)[1]. For concreteness we use the following assumptions:

1. The jet is highly relativistic, $\gamma \gg 1$.

2. The initial photons are soft (i.e., $h\nu_0 \ll 511$keV).

[1]An alternative situation is a continuous photon source illuminating a very short, collimated, lump of electrons, i.e. a relativistic "cannon ball" moving in a radiation field.

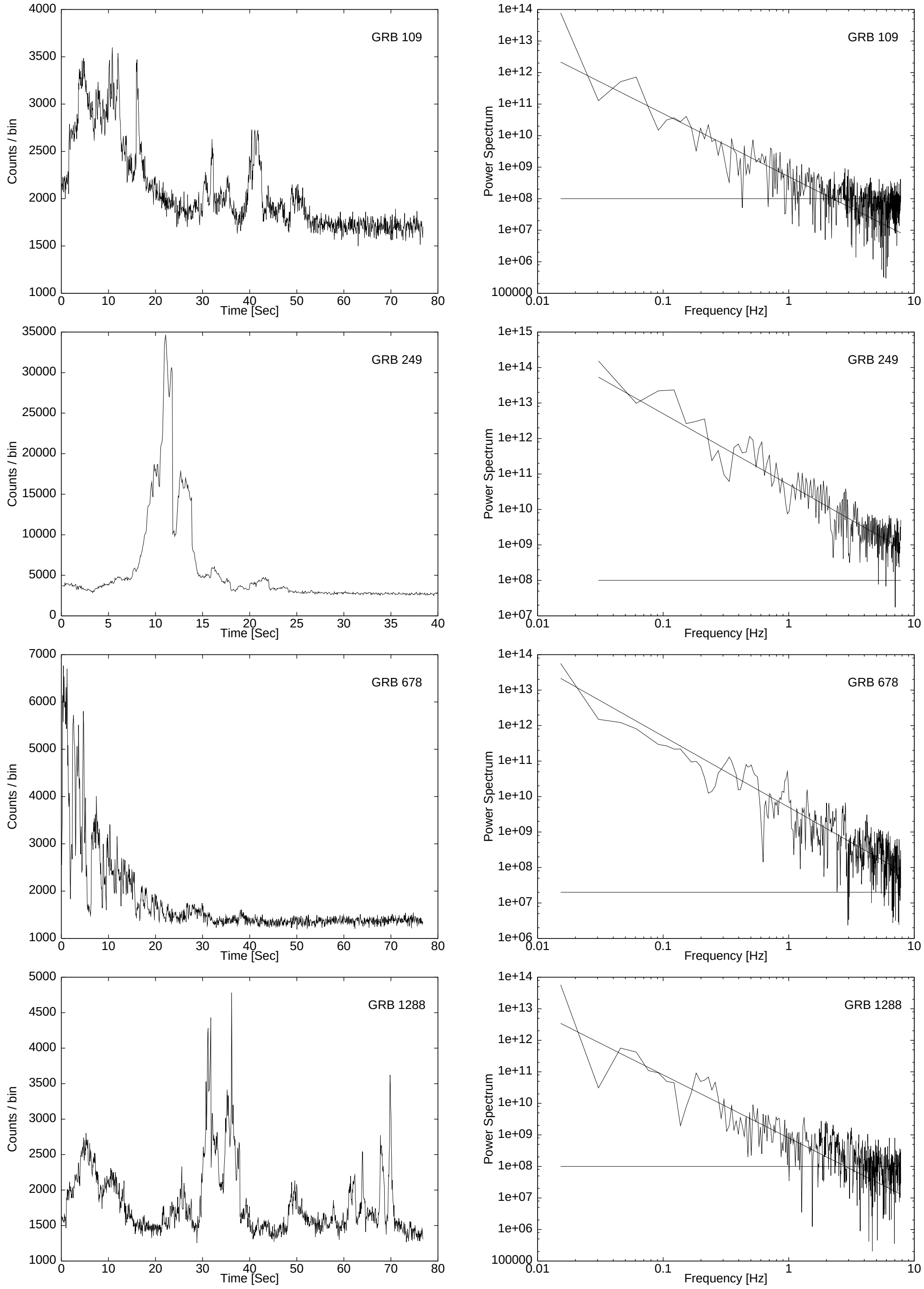

Fig. 3.— The light curves and the corresponding power spectra of four GRBs with complex light curves, arbitrarily selected from the BATSE 1B catalogue. The straight lines represent an $\omega^{-2}$ power spectrum. The horizontal lines represent the power spectrum of the observed background after the GRBs fell well below the background.

3. The photon source size is small compared with the interaction length, i.e., $\phi \ll 1$ where $\phi$ is the incident angle of the scattered photons.

4. The product $\gamma\phi$ satisfies $\gamma\phi \gg 1$ .

5. The total fluence needed for detection of the GRB *is independent* of the duration of the GRB. (This assumption is correct for burst durations which are shorter than the duration of the trigger bin. If the burst is smooth, the detection probability will decrease thereafter. On the other hand, if the bursts have sparse spikes, the detection probability will start to decrease drastically only when the spikes themselves are of the order of the duration of the trigger bin. The actual case is more complicated, and will be elaborated further on.)

The frequency of the initial photons in the observer rest frame, $\nu_0$, and their frequency in the jet rest frame, $\nu^*$, are related through:

$$\nu^* = \gamma(1 - \beta\cos\phi)\nu_0. \tag{4}$$

Note that $\phi$ which is the angle between the photon's momentum and electron's axis (in the observers frame) cannot be equal to 0, which then gives no "acceleration" of the photons. The relation between $\phi$ and $\phi^*$ (the asterisk stands for the jet's rest frame) is:

$$\sin\phi^* = \frac{\sin\phi}{\gamma(1 - \beta\cos\phi)}. \tag{5}$$

However, according to our assumptions, $\phi \ll 1$ and $\gamma\phi \gg 1$, hence, $\sin\phi \approx \phi$, $\cos\phi \approx 1 - \phi^2/2$, $\beta \approx 1 - \gamma^{-2}/2$ and

$$\sin\phi^* \approx \phi^* \approx \frac{2}{\gamma\phi} \ll 1. \tag{6}$$

The Compton cross section in the electron rest frame changes slowly with angle and does not diverge at small angles. Therefore, $\phi^*$ is very small compared with the typical values of $\theta^*$ (the scattering angle in the electron rest frame) and can be taken as *zero*. In addition:

$$\nu^* = \gamma(1 - \beta\cos\phi)\nu_0 \approx \frac{\gamma\phi^2}{2}\nu_0. \tag{7}$$

For $h\nu^* \ll 511keV$ Compton scattering does not change the photons' energy in the electron's rest frame. Consequently, photons that satisfy

$$h\nu_0\frac{\gamma\phi^2}{2} \ll 511keV \tag{8}$$

will not be boosted by Compton scattering.

If $\nu$ and $\theta$ are the frequency and angle of the scattered photon relative to the jet axis in the observer frame, then,

$$\nu = \frac{\gamma^2\phi^2}{2}(1 + \beta\cos\theta^*)\nu_0. \tag{9}$$

The differential cross sections in the electron rest frame for Compton scattering of soft photons, after averaging over initial polarizations and summing over final polarizations, is given by

$$\frac{d\bar{\sigma}}{d\Omega^*} = \frac{\alpha}{2m_e^2}(2 - \sin^2\theta^*). \tag{10}$$

The scattering angles in the electron and observer rest frames are related through:

$$\sin\theta^* = \frac{\sin\theta}{\gamma(1-\beta\cos\theta)} \approx \frac{2}{\gamma\theta + \frac{1}{\gamma\theta}}, \tag{11}$$

where we used the condition $\theta, \gamma^{-1} \ll 1$. As a consequence, one obtains the following photon fluence (photons ster$^{-1}$) in the electron's frame:

$$\frac{d\Phi^*(\theta^*)}{d\Omega^*} = \Phi_0(2-\sin^2\theta^*) \approx \Phi_0\left(2 - \frac{4}{(\gamma\theta+\frac{1}{\gamma\theta})^2}\right). \tag{12}$$

Conservation of photon number yields

$$\frac{d\Phi^*}{d\Omega^*}\sin\theta^* = \frac{d\Phi}{d\Omega}\sin\theta, \tag{13}$$

hence,

$$\frac{d\Phi}{d\Omega} = \frac{d\Phi^*}{d\Omega^*}\frac{\sin\theta^*}{\sin\theta}. \tag{14}$$

If the GRB detectors have a constant response in a given spectral region and if the GRBs have a power law spectrum with a spectral index $\alpha$ (typically between $-2$ to $-1$) then one can relate the total energy received in the detectors to the photon flux. The energy fluence per unit solid angle, $I(\theta)$ is given then by

$$I(\theta) = I_0\left(\frac{\nu}{\nu_0}\right)^{-(\alpha+1)}\frac{d\Phi(\theta)}{d\Omega} \tag{15}$$

with $I_0$ being a normalization constant.

The relation between the duration $t$ and the scattering angle $\theta$ for a GRB can be read from figure 5. If $l_0$ is the typical jet length from where the $\gamma$ rays are produced, the path difference between two light rays scattered from the two ends of the jet is given by

$$\Delta l = l_0(1-\cos\theta). \tag{16}$$

The corresponding time difference, which according to our model is the burst duration, is given for $\theta \ll 1$ by

$$t = \frac{l_0}{c}(1-\cos\theta) \approx \frac{l_0}{2c}\theta^2. \tag{17}$$

Hence, we have the following relations:

$$dt = \frac{l_0}{c}\sin\theta d\theta;\ \frac{d\theta}{dt} \approx \frac{c}{l_0\theta}. \tag{18}$$

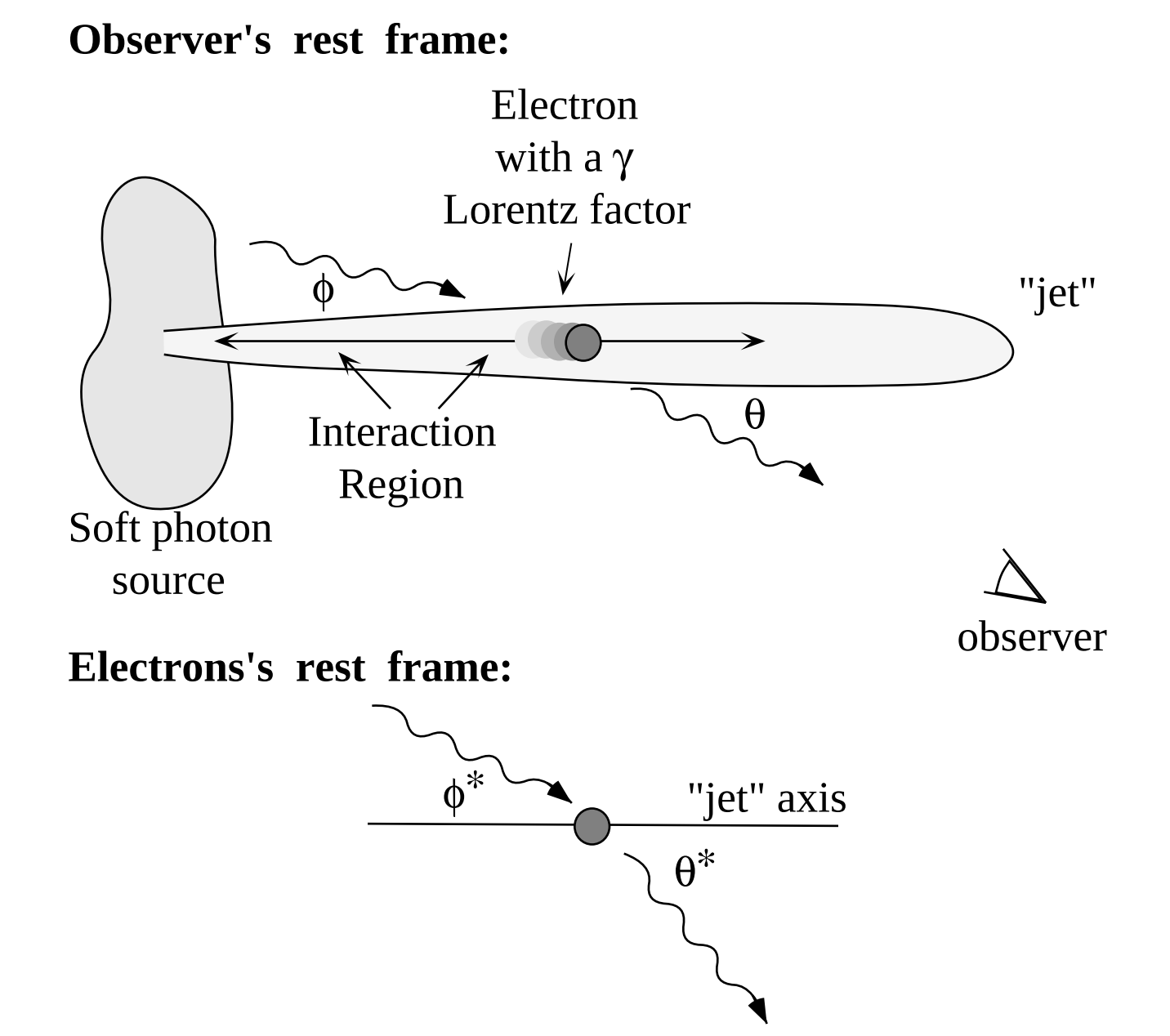

Fig. 4.— A schematic description of a jet model for production of GRBs. Soft photons are Compton scattered from a relativistic jet. The angles $\phi$ and $\theta$ are the photon angles relative to the jet in the observer frame before and after the scattering, respectively. The asterisks denote angles in the jet (electrons) rest frame.

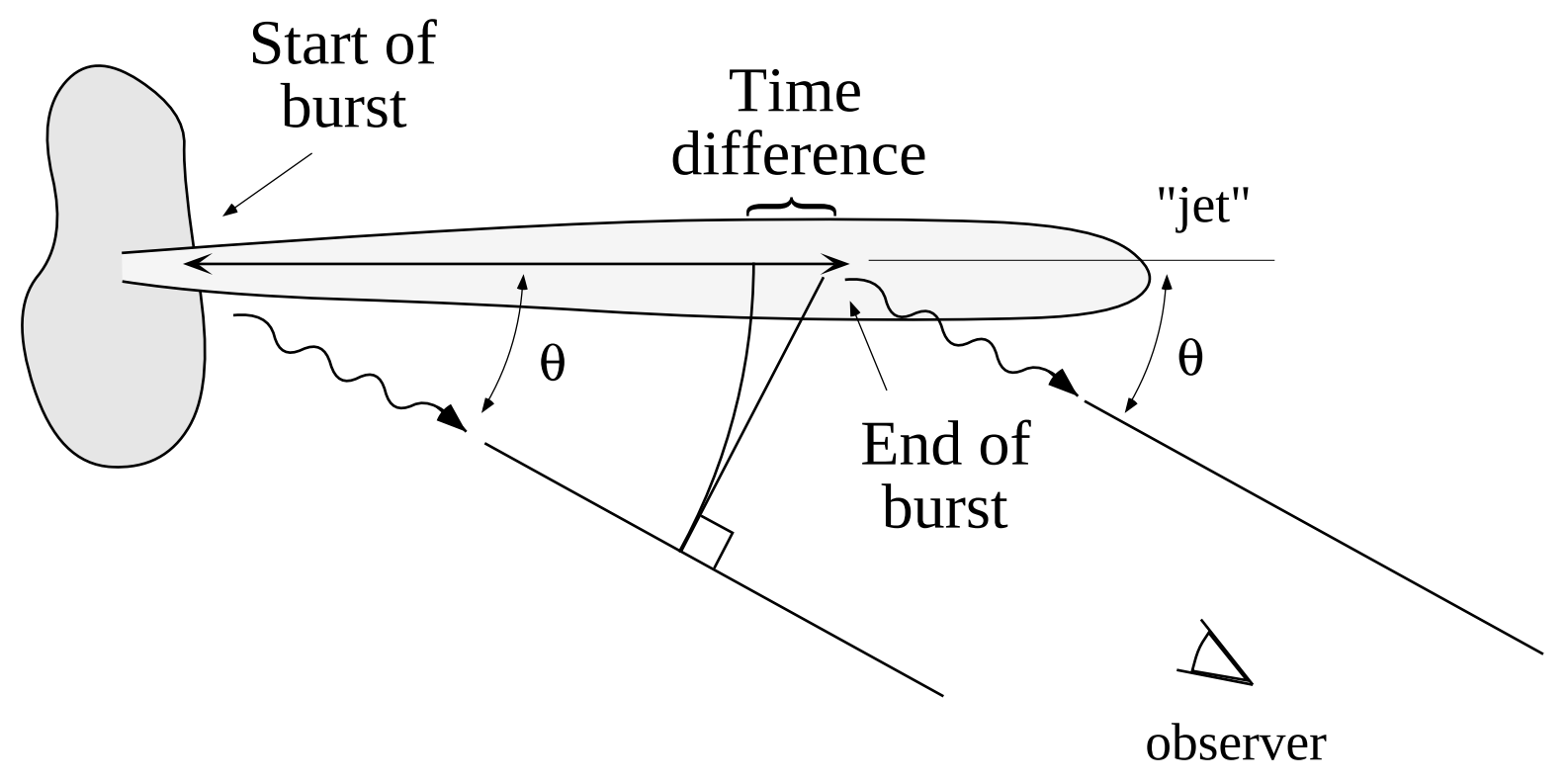

Fig. 5.— Formation of different burst durations in a jet model by a geometrical effect. The burst duration reflects the difference in the path length of photons scattered from the two opposite ends of the jet.

### 3.2. Approximate Duration Histogram from the Simplest Model

Let us first employ simplifying assumptions in order to derive an approximate result and some immediate conclusions concerning the duration histogram of GRBs. More realistic estimates and more accurate results are derived in the next subsection.

In order to relate $I(\theta)$ to the number of observed GRBs we will first assume that all GRBs with a fluence larger than a minimum fluence, $F_{min}$ (erg cm$^{-2}$), are detected. This, of course, is not accurate, and will be corrected later. The largest distance from which a GRB with this threshold fluence can be detected is

$$R_{max} = \sqrt{\frac{I(\theta)}{4\pi F_{min}}}. \tag{19}$$

For an Euclidean space, the number of GRBs inside a sphere of radius $R_{max}$ is proportional to $R_{max}^3$. The number of bursts with their jets oriented within an angular interval $d\theta$ along the line of sight $d\theta$ which are detected is thus given by:

$$\frac{dN(R_{max})}{d\theta} = N_0 R_{max}^3 \sin = N_0 \left(\frac{I(\theta)}{4\pi F_{min}}\right)^{\frac{3}{2}} \sin\theta\ , \tag{20}$$

where $N_0$ is a normalization constant. Using Eq. 18 one can rewrite the last equation as

$$\frac{dN}{dt} = \frac{N_0 c}{l_0}\left(\frac{I(\theta)}{4\pi F_{min}}\right)^{\frac{3}{2}}\ . \tag{21}$$

If one inserts the approximate value for $I(\theta)$ (for a highly relativistic jet),

$$I(\theta) = I_1\left(2 - \frac{4}{\left(\gamma\theta + \frac{1}{\gamma\theta}\right)^2}\right)\left(\gamma\theta^2 + \frac{1}{\gamma}\right)^{-1}\left(\frac{2}{1+\gamma^2\theta^2}\right)^{-(\alpha+1)}, \tag{22}$$

one obtains for $dN/d\log t = tdN/dt$ (number of events seen per unit $\log t$), the prediction is:

$$\frac{dN}{d\log t} = \frac{N_0 c}{l_0}\left(\frac{I_1}{4\pi F_{min}}\right)^{\frac{3}{2}}\left\{\frac{\left(2 - \frac{4}{\left(\gamma\theta+\frac{1}{\gamma\theta}\right)^2}\right)}{\left(\gamma\theta^2 + \frac{1}{\gamma}\right)}\left(\frac{2}{1+\gamma^2\theta^2}\right)^{-(\alpha+1)}\right\}^{\frac{3}{2}}\ . \tag{23}$$

Fig.6 compares this prediction for $dN/d\log t$ and the histogram of the $T90$ durations of the GRBs in the BATSE 1B catalogue. The fit was obtained with only two parameters - $\gamma$ and the normalization constant for the two populations of GRBs. This model shows its basic feature, namely, the very wide histogram of durations for one population (three orders of magnitude in duration!) can be approximately reproduced. In addition, we find that the shorter bursts of each population have a brighter intrinsic fluence, hence, two bursts of different durations but of the same fluence, do not have the same distance, with the shorter coming from a further distance.

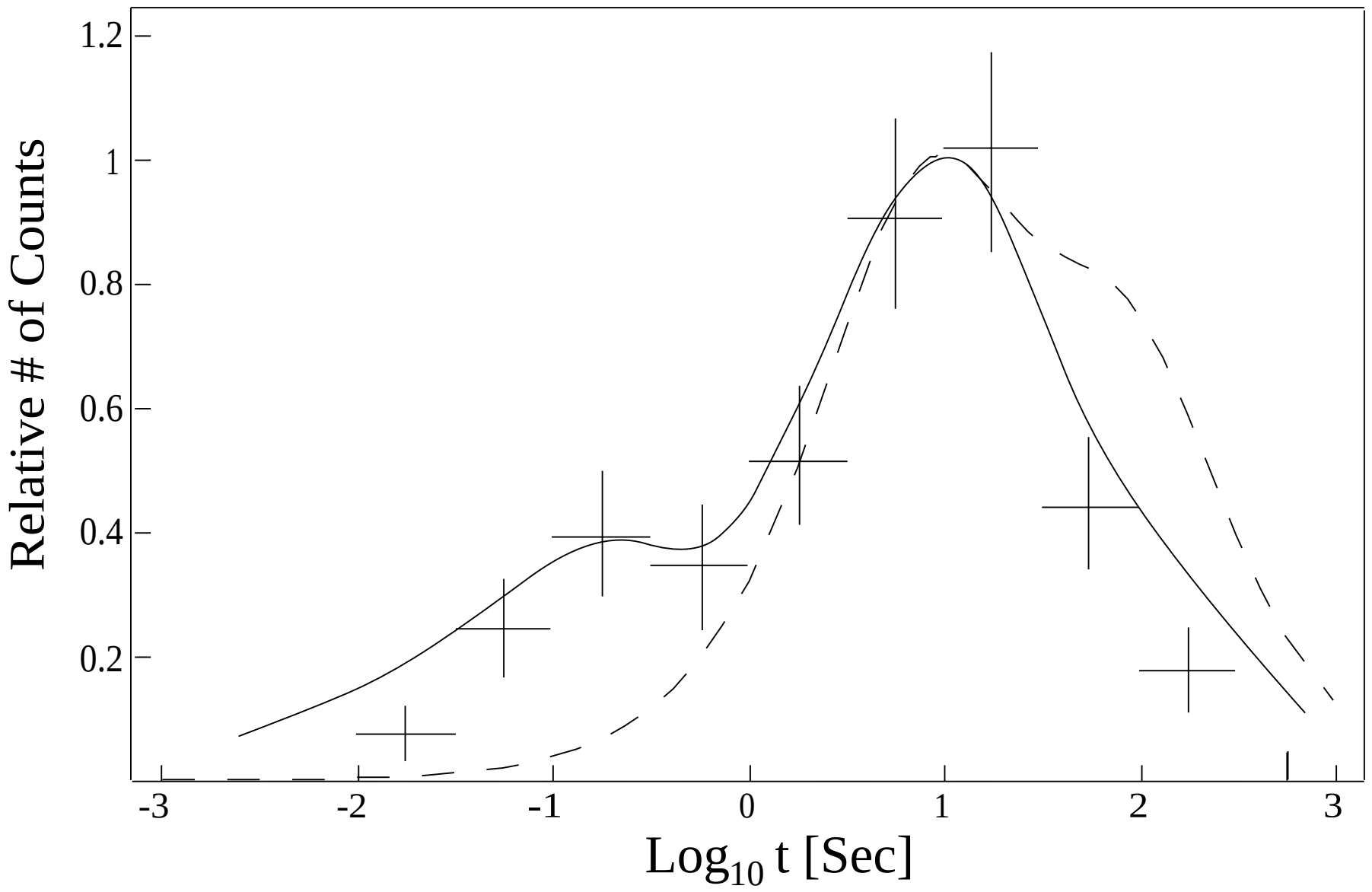

Fig. 6.— Comparison between the duration histogram of GRBs from the BATSE 1B catalogue and the prediction of the simplest "jet" model which assumes one population, a constant "fluence trigger", and no cosmological effects (dashed line). Adding an additional population, and a simple fluence threshold criterion as summarized by Eq. 24, improves the agreement between the observations and the predicted histogram (full line). The remaining discrepancies may be removed by introducing more realistic threshold criteria, cosmological effects, and/or by spreading the values of the Lorentz factor $\gamma$.

### 3.3. Duration Histogram from Realistic Considerations

We have shown in the previous section that the simplest jet model is able to produce a fairly good fit to the observed duration histogram of GRBs even before including cosmological effects and the effects of experimental biasing of the detection threshold. In this section we improve the agreement between the model prediction and the observed histogram by including these effects. They can be included analytically, but this involves lengthy mathematics and nontransparent equations. Rather, we prefer to present the results of numerical "Monte Carlo" simulations of the effects, which are simple, flexible and allow the study of additional effects.

**Threshold Biasing** - As was mentioned before, it is very complicated to cast the $C_{max}/C_{min}$ (the ratio between the peak count rate and the threshold count rate) criterion into an effective fluence criterion, which is needed in order to compare between the GRBs duration histogram predicted by the jet model and the duration histogram of observed GRBs. Moreover, the sought relation requires knowledge of the exact average burst profile which is not well known. Therefore, the following procedure was adopted. Instead of using all the data we used only part of the data that passes a simple explicit fluence criterion of our own choice. The criterion was chosen in such a way that on one hand it does not filter in bursts that do not pass the $C_{max}/C_{min}$ criterion, and on the other hand it does not reduce significantly the statistics of the data set. The criterion chosen for $F_{min}$ is

$$F_{min} = F_0\sqrt{t} \; . \tag{24}$$

$F_0$ was chosen to be $1.0 \times 10^{-7}$ erg cm$^{-2}$ with $t$ measured in seconds. The effect of this threshold criterion on the BATSE 1B catalogue is shown in Fig. 7. Indeed, most of the detected GRBs are surviving this criterion.

**Cosmological Effects** - The following two relations were used in the Monte Carlo simulations of cosmological GRBs: First, the relation between the redshift $z$ and the luminosity distance $D_L$ to an object,

$$D_L = \frac{R_H}{q_0^2}\left[1 - q_0 + q_0 z + (q_0 - 1)\sqrt{2q_0 z + 1}\right] \; , \tag{25}$$

where $R_H = c/H_0$ is the Hubble radius and $q_0$ is the decelaration parameter. The second relation that was used is the number of cosmological objects in a $dz$ interval assuming no evolution:

$$\frac{dN}{dz} = \frac{n_0 R_H^3}{(1+z)^3}\frac{[q_0 z + (q_0 - 1)(\sqrt{1 + 2q_0 z} - 1)]}{q_0^4\sqrt{1 + 2q_0 z}} \; , \tag{26}$$

where $n_0$, is a normalization constant. Unlike the number density of galaxies per comoving volume, for example, which does not change much for $z$ values that are smaller than $z$ for the galactic formation epoch, the number density of bursts could depend on $z$, since the formation rate of the progenitors could depend on $z$.

**Simulation** - All the above ingredients were incorporated in a Monte Carlo simulation to find the different distributions. Bursts were simulated by randomly choosing different $z$ values and different values of $\theta$. For each burst, the intrinsic intensity was found and the observed intensity was calculated. Then it was checked whether the burst passes the threshold criterion. From accumulating the simulated bursts that passed the threshold criterion a duration histogram was finally drawn. More details are described below.

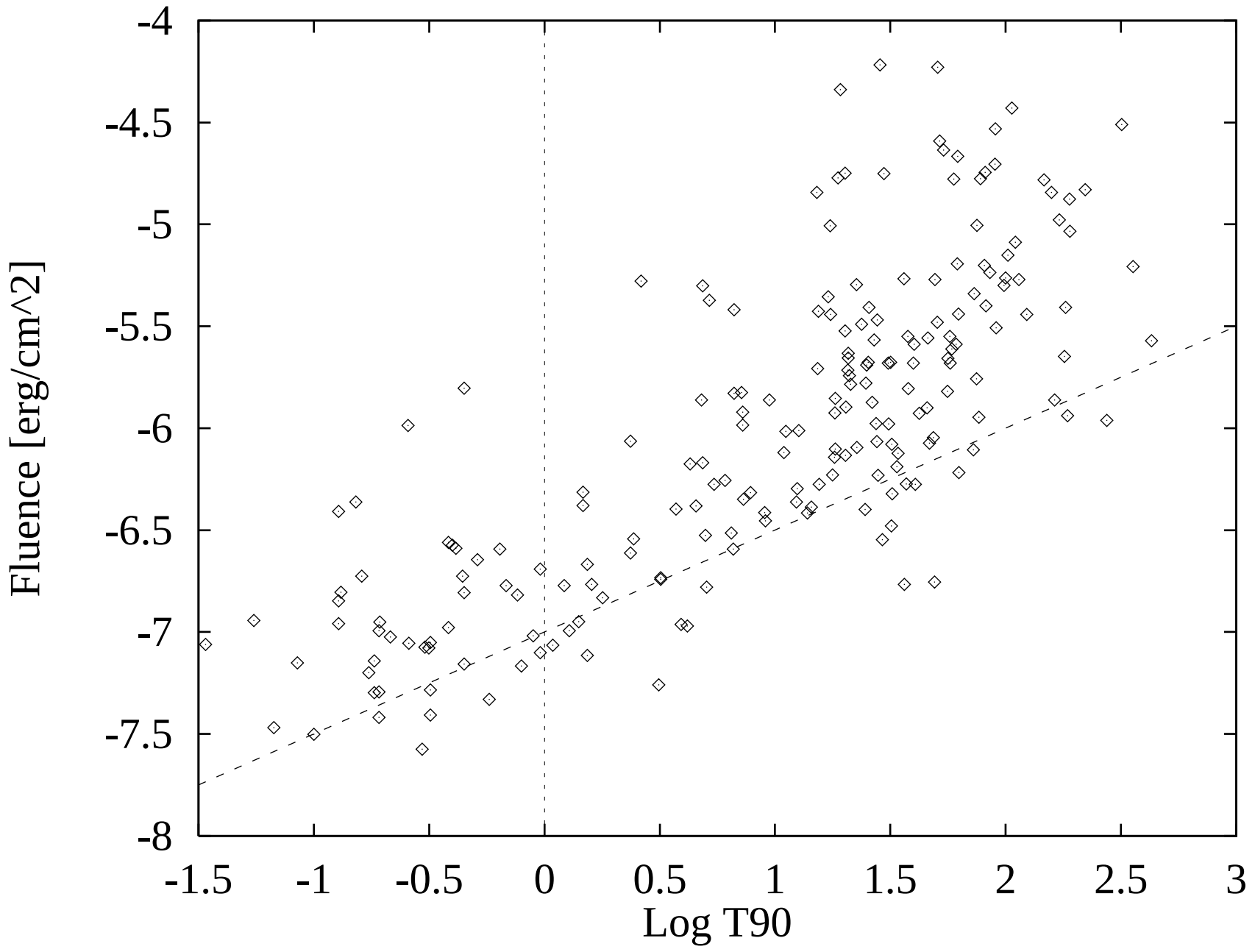

Fig. 7.— The fluence threshold (dotted line) used for comparing the predicted duration histogram of GRBs and the duration histogram of the GRBs in the BATSE 1B catalogue (doted boxes).

## 4. Comparison with Observations

### 4.1. Duration Histograms

The duration histogram of GRBs predicted by the jet model was calculated for a cosmic density parameter and a cosmological constant of $\Omega = 2q_0 = 1$ and $\Lambda = 0$, respectively. The average spectral index of gamma ray bursts was taken as $\alpha = -1.5$ (the results are not sensitive to the exact values of these parameters). The normalizations and the peak positions of the populations of short and long duration GRBs were fitted directly to the experimental data. The predicted histogram is compared in Fig. 8 with the histogram obtained from the durations of the GRBs in the BATSE 1B catalogue. As can be seen from Fig. 8 the agreement between the theory and the observations is very good ( A $\chi^2$ of 1.05 for 3 degrees of freedom was obtained yielding a confidence level of 0.80).

### 4.2. Relative Distances

As was already pointed out, the jet model predicts that the shorter bursts in a burst population generated by the same jets are further away than the longer bursts produced by the same jets. To verify this we divided the long duration population of GRBs in the BATSE 1B catalogue into two groups: one group between 1s and 10s, and the second group between 10s and 100s (giving almost the same number of bursts in each group). For each group we plotted the number of bursts with $V_{max}/V_{min} \equiv (C_{max}/C_{min})^{3/2}$ smaller than a given value. As can be seen from Fig.9 the "break away" from Euclidean space occurs at different values of $C_{max}/C_{min}$ for the two groups with the longer duration bursts breaking away at a value approximately 3 times smaller than that for the shorter duration bursts. The break away points can be translated into fluences using the Monte Carlo simulation for estimating the effect of the threshold on the extracted fluences. The observed ratio of the fluences at the break away points is $0.35 \pm 0.15$ in good agreement with the value $0.3 \pm 0.1$ obtained from the Monte Carlo simulation of jet produced GRBs.

### 4.3. Energy Spectrum

From Eq. 9 it follows that the energy spectrum of a GRBs depend on the energy spectrum of the electrons in the jet, the energy spectrum of the soft photon source and the source intensity as function of distance from the jet axis. If the light source that illuminates the jet is monochromatic then the energy spectrum of the scattered light depends on the energy distribution of the electrons in the jet. In particular, if electrons in the jet have a power law spectrum $dn_e/dE \sim E_e^{-n}$ they will generate a power-law spectrum of Compton scattered photons $dn_\gamma/dE \sim E_\gamma^{-(n+1)/2}$. The power-law index may depend on the distance along the jet and may give rise to time-dependent spectra, e.g., to harder spectra (but smaller intensities) towards the end of the GRBs if electrons are accelerated in the jets and to softer spectra if electrons are slowed down along the jets.

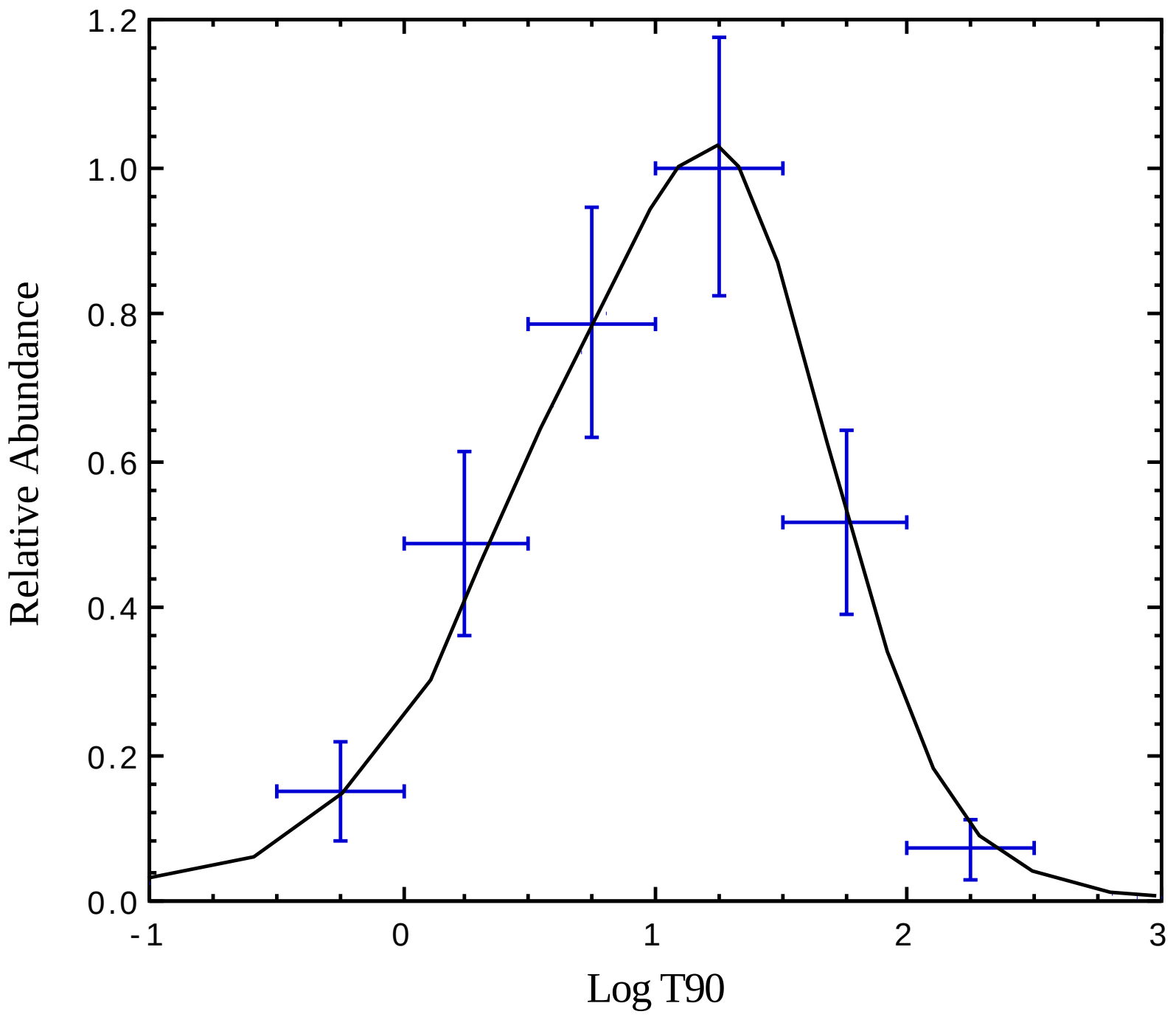

Fig. 8.— Comparison between the duration histogram of GRBs derived from the Monte-Carlo simulation of jet produced GRBs (line) and the duration histogram for the "long duration" GRBs in the BATSE 1B catalogue (crosses).

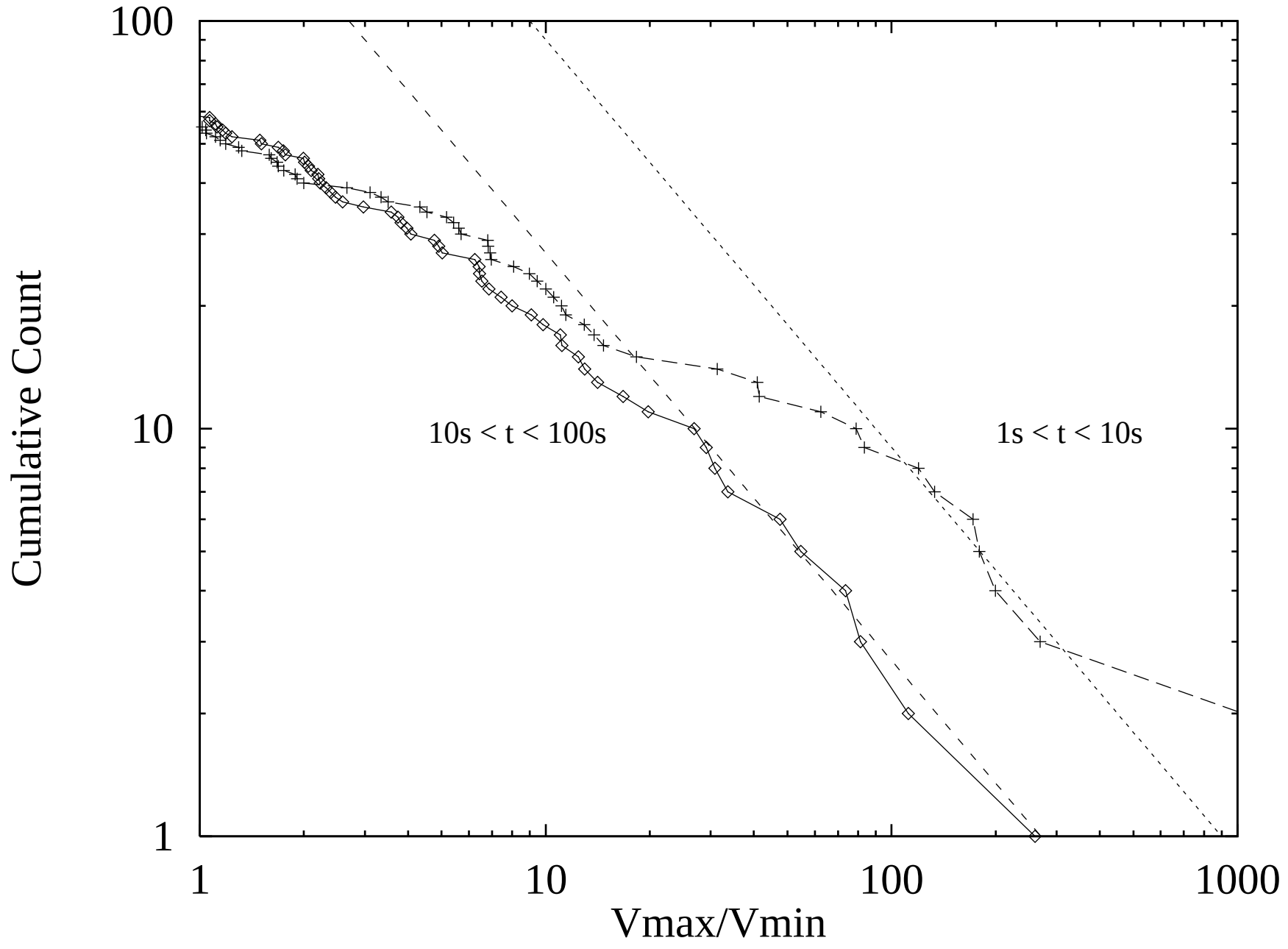

Fig. 9.— The cumulative number of bursts as function of $V_{max}/V_{min} \equiv (C_{max}/C_{min})^{3/2}$, for two GRBs populations, one with burst durations between 1s and 10s and the other with burst durations between 10s and 100s.

### 4.4. Light Curves

High resolution radio observations (VLA and VLBI) of AGN jets show that the jets are not continuous but made of "knots". This is demonstrated in Figs. 10,11 for the jet in NGC 6251 (Perley, Bridle and Willis 1984). Such a substructure which seems to be common to AGN jets may be a universal feature of highly relativistic jets due to their production mechanism or due to magnetohydrodynamic instabilities in the jets. The power spectrum of this segmented distribution along the jet axis seems to be given by $\sim \omega^{-2}$ (see Fig. 11b). If the segmented jet structure with such a power spectrum is typical also to minijets, then inverse Compton scattering from such jets will give rise to multipeak gamma ray bursts that look very complex but have a rather simple, $\sim \omega^{-2}$, power spectra, which we found (see, e.g., Fig. 3) to be a general feature of GRBs. (A uniform jet structure can also give rise to a multipeak emission provided that it is illuminated by a light source with a multipeak light curve or it is propagating in a highly non uniform radiation field). We defer, however, the detailed study of the multipeak structure of GRBs light curves to future publications.

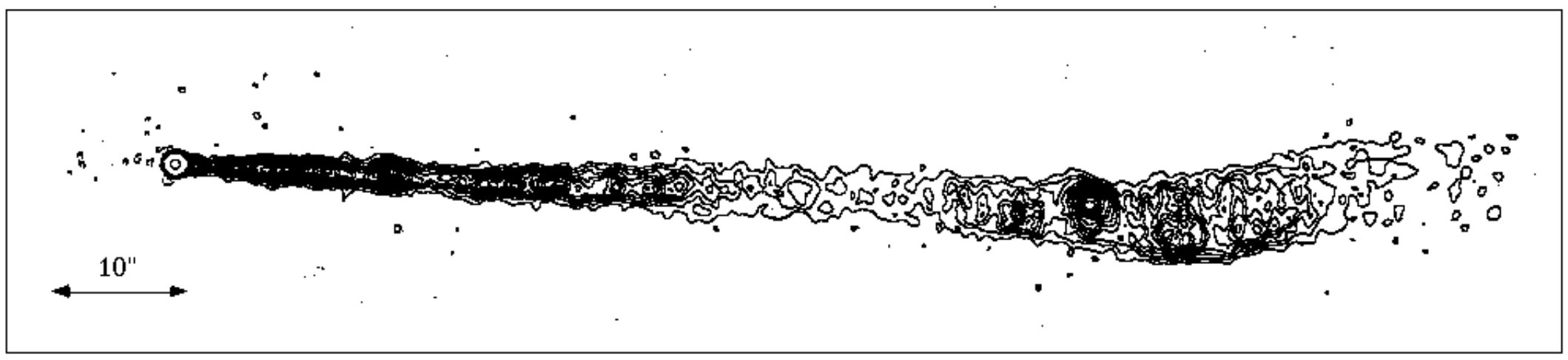

Fig. 10.— A VLA radio map of the jet in NGC 6251 obtained by Perley, Bridle and Willis 1984.

## 5. Predictions

In addition to explaining observed features of GRBs, the jet model of GRBs production can be used to make several additional predictions. One interesting prediction is that the observed radiation should be polarized. A second interesting prediction is a relation between the length of the jet and the value of the electrons' $\gamma$-factor. A third prediction is an estimate of the typical photon energy released in a GRB.

### 5.1. Polarization of GRBs

Quite generally, Compton scattered radiation from a highly relativistic jet is partially polarized with the direction of the electric field perpendicular to the jet axis. The degree of polarization as a function of scattering angle is given by the following simplified expression for *soft* photons in the electron's rest frame, (e.g. McMaster 1961, Skibo et al. 1994):

$$\Pi^*(\theta^*) = \frac{\sqrt{Q^2+U^2+V^2}}{I} = \frac{\sin^2\theta^*}{2-\sin^2\theta^*}, \tag{27}$$

and I,Q,U and V are the Stoke's parameters of the scattered radiation. Since the degree of polarization is a Lorentz invariant quantity, i.e., $\Pi^*(\theta^*) = \Pi(\theta)$, one finds:

$$\Pi(\theta) = \frac{\sin^2\theta^*}{2-\sin^2\theta^*} = \frac{2}{\left(\gamma\theta+\frac{1}{\gamma\theta}\right)^2 - 2}\ . \tag{28}$$

Using Eq. 17, one can predict the average polarization of a GRB as function of its duration.

Figure 12 shows the predicted polarizations of long duration GRBs as function of their duration after inserting relation 17 between $\theta$ and $t$. As can be seen, the polarization has a maximum for $t$ of the order of 5 Seconds. At maximum, the gamma rays are almost completely polarized.

Figure 13 shows a histogram of expected polarization, for a $\gamma$-ray burst population passing the threshold detailed in section 3.3. When measuring the effective duration of the bursts, there is a large error arising from the high intrinsic variability of the bursts. In addition, the polarization is not a unique value for a given duration, even without dispersion, because the bursts are generated at different red-shifts. Hence, if one wants to compare the predicted average polarizations with future observations Fig. 12 may not be useful enough. A histogram of polarizations may then be more useful for comparisons.

### 5.2. Jet Length and Lorentz $\gamma$ Factor

We shall now proceed to evaluate the length of the jet from where 90% of the counts arise. For the GRBs from the BATSE 1B catalogue the maximum of the histogram of $T90$ durations occurs at $t_{max} \approx 15$ sec (this is not the place where the maximum occurs in the BATSE data, but rather the place where the maximum occurs after the data are passed through our artificial threshold criterion describe by eq. 24). Hence, from eq. 17 it follows that

$$t_{max} \approx \frac{l_0 \theta_{max}^2}{2c} \approx 15 \text{ sec.} \tag{29}$$

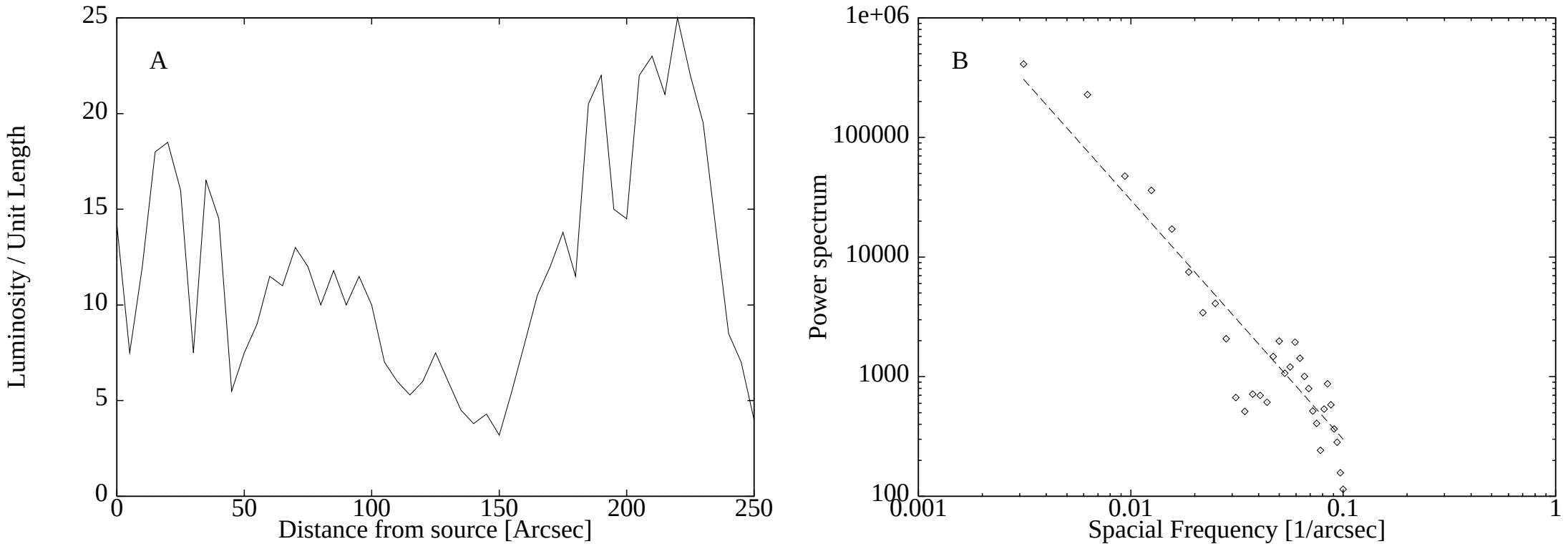

Fig. 11.— (a) The intensity of radio emission as function of distance along the jet axis for the jet in NGC 6251 obtained by Perley, Bridle and Willis 1984. (b) the power spectrum of the distribution of radio emission along the jet axis for the jet. The straight line represents an $\omega^{-2}$ power spectrum.

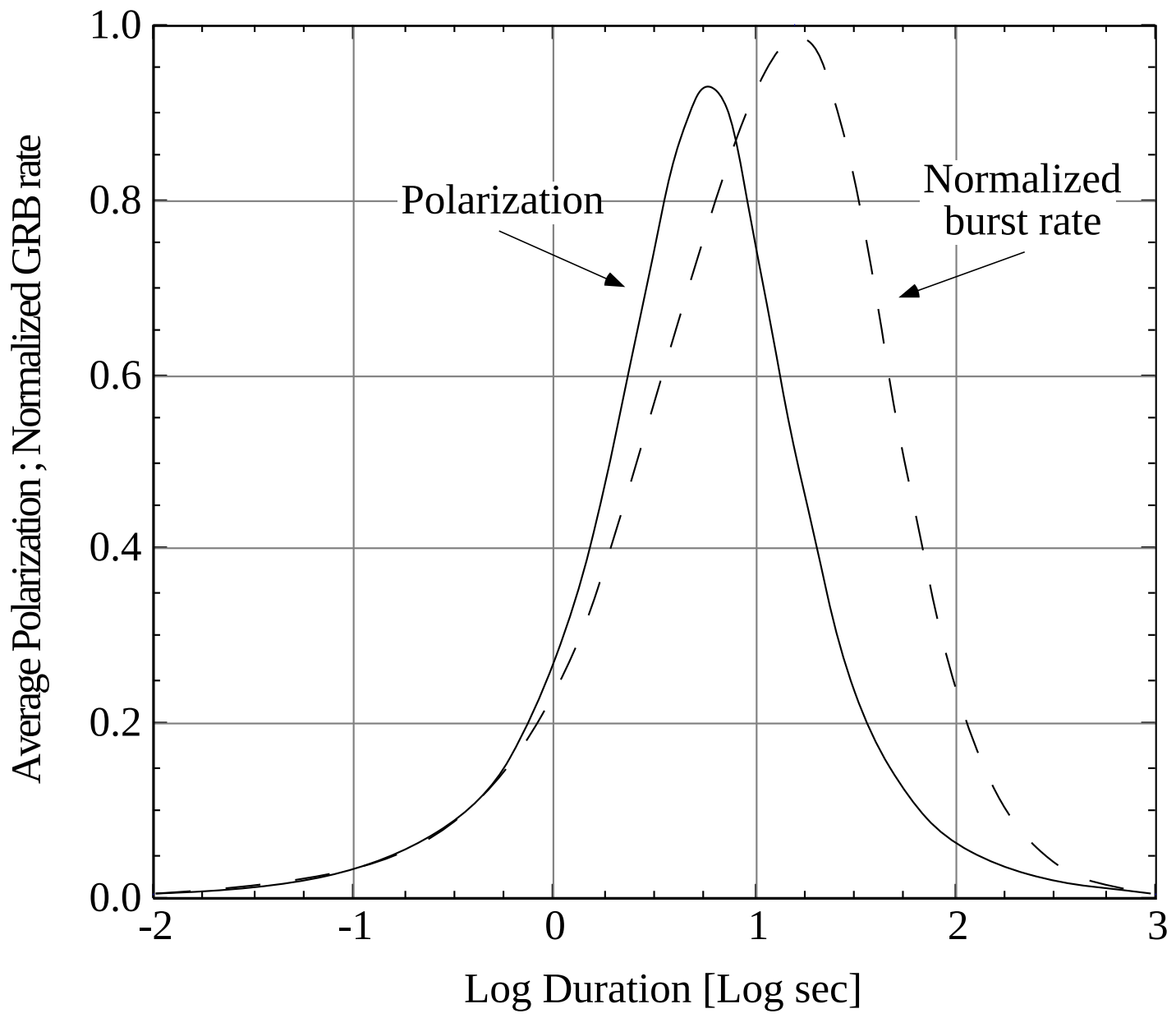

Fig. 12.— Average polarization of long duration GRBs as a function of their duration. The maximum average polarization occur for durations of about 5 seconds where the average polarization is nearly complete. The normalized distribution of GRBs durations for the same GRBs population is also shown for comparison.

From the Monte Carlo simulations we found $\theta_{max}\gamma \approx 0.6$, hence

$$l_0 \approx 6c\gamma^2 t_{max} \approx 95\gamma^2 c \text{ sec,} \tag{30}$$

i.e., $l_0 \approx 3 \times 10^{-2}(\gamma/100)^2$ light years.

### 5.3. Estimate of the Intrinsic GRB Luminosity

The simple jet model allows us to calculate the total photon luminosity, if the distance scale of the sources and the Lorentz $\gamma$ factor of the jets are given. The distance scale can be found from the $\chi^2$ fit of the model to the observations but the error is rather large. The $\gamma$ factor however, cannot be found from the model.

Since we are interested only in an approximate estimate, we assume for simplicity that the total emitted energy is approximately the energy emitted in our direction multiplied by the ratio between the total number of emitted photons and the number of photons emitted in our direction. If $\Phi$ is the time integrated photon flux, and $R$ is the distance to the gamma ray burster, then the measured fluence is

$$F = F_0 \frac{\partial \Phi(\theta_0)}{\partial \Omega} \frac{1}{R^2}, \tag{31}$$

where $F_0$ is a normalization constant (relating photon number and fluence). Let us define a function $f(\theta)$ and a normalization constant $\Phi_0$, as in Eq. 12,

$$\frac{\partial \Phi(\theta)}{\partial \Omega} = \Phi_0 f(\theta), \tag{32}$$

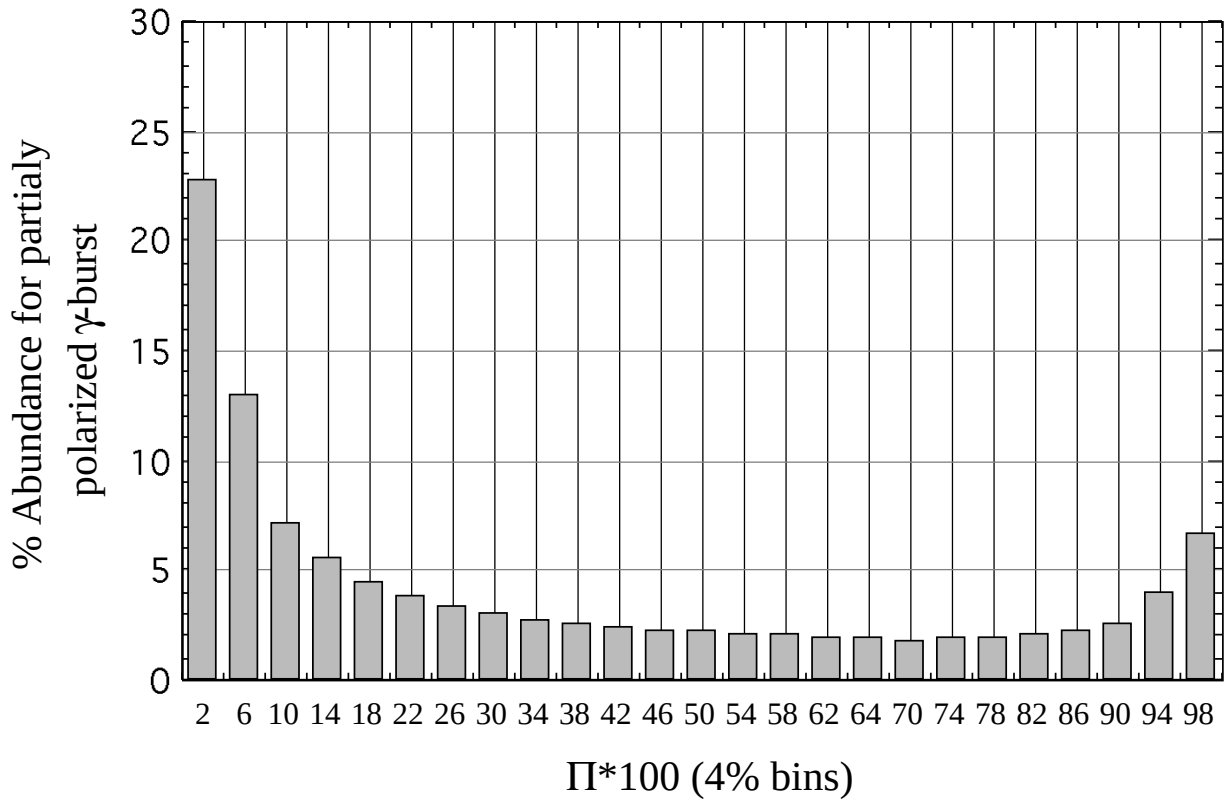

Fig. 13.— Histogram of the expected polarizations (4% bins) calculated with the artificial threshold described in 3.3.

with $f(\theta)$ being

$$f(\theta) = \left(2 - \frac{4}{\left(\gamma\theta + \frac{1}{\gamma\theta}\right)^2}\right)\left(\gamma\theta^2 + \frac{1}{\gamma}\right)^{-1}. \tag{33}$$

We therefore have

$$\Phi_0 = \frac{R^2 F}{F_0 f(\theta_0)}. \tag{34}$$

The total emitted energy is thus

$$E_{tot} = 2 \times \int \frac{\partial \Phi(\theta)}{\partial \Omega} d\Omega = \frac{2R^2 F}{f(\theta_0)} \int f(\theta) d\Omega, \tag{35}$$

where

$$\int f(\theta) d\Omega = \int f^*(\theta^*) d\Omega^*. \tag{36}$$

One can calculate the integral in the electrons' rest frame and obtain

$$\int f^*(\theta^*) d\Omega^* = \int_0^\pi 2\pi \sin\theta^* (2 - \sin^2\theta^*) d\theta^* = \frac{16\pi}{3}, \tag{37}$$

which leads to

$$E_{tot} = \frac{32\pi R^2 F}{3\gamma} \frac{x^2 + 1}{2 - \frac{4}{(x+1/x)^2}}, \tag{38}$$

with $x = \gamma\theta$, instead of $E_{tot} = 4\pi R^2 F$. for a spherically symmetric fireball. If one chooses $x \approx 10^{-0.25}$ (i.e. $t \approx 10$ sec), one finds that $z_{max} \approx 2$ yields for $\Omega = 1$ a distance $R \approx 7.6 h^{-1}\ Gpc$ where h is the Hubble constant in units of 100 $km\ s^{-1} Mpc^{-1}$. From Fig.7 one finds that $F \approx 10^{-6.5}$, yielding:

$$E_{tot} \approx \frac{6 \times 10^{51}}{\gamma h^2}\ ergs. \tag{39}$$

Since we have used the fluence in the BATSE bandwidth this total energy is approximately the total energy emitted in this bandwidth.

## 6. Discussion and Conclusions

In this paper we explored the possibility that inverse Compton scattering of light from highly relativistic transient jets may provide a universal mechanism for the production of GRBs.

Highly relativistic transient jets may be produced along the axis of an accretion disk formed around stellar black holes or neutron stars in BH-NS and NS-NS mergers and/or in accretion induced collapse of magnetized white dwarfs or neutron stars in close binary systems. Such events in external galaxies may produce the cosmological GRBs. Highly relativistic transient jets may also be formed by single old magnetized neutron stars in an extended Galactic halo. They may be illuminated by a short burst of photons following a phase transition or a starquake in the neutron star. Such events may produce a local population of GRBs. These two distinct sources of highly relativistic jets may explain the existence of the two distinct populations of GRBs, the long duration population ($> 1$ *sec*) and the short duration ($< 1$ *sec*) population, respectively.

Many similarities between gamma ray emission from AGNs and GRBs do suggest that GRBs are produced by highly relativistic transient jets illuminated by a light source. Inverse Compton scattering of photons from the light source by highly relativistic electrons in the jet boosts their energies to gamma ray energies and beams them in a narrow cone along the jet. Indeed, in this paper we have shown, using a simple jet model, that jet production of GRBs by inverse Compton scattering can explain the striking correlations that exist between various temporal features of GRBs, their duration histogram and their power-law high energy spectrum. Instabilities in the jets and/or in the illuminating sources were proposed as the source of the diversity and complexity of GRBs light curves. Such a mechanism can account for the simple power spectrum ($\sim \omega^{-2}$) that we found for the complex multipeak GRBs light curves, and will be further discussed in future publications. Some predictions were made, including the expected polarization of the gamma-rays in the bursts and a polarization histogram for GRBs. These are independent of the exact knowledge of the Lorenz $\gamma$ factors of the jets. In conclusion, jet production of GRBs seem to be able to explain many of the observed features of GRBs and to avoid many of the problems encountered by the relativistic expanding fireball model. This promising direction of investigation should be pursued.

## Acknowledgements

This research has made use of data obtained through the Compton Observatory Science Support Center GOF account, provided by the NASA-Goddard Space Flight Center. The authors would like also to thank Amos Ori for the fruitful discussions.